\documentclass[a4paper,12pt]{article}

\def\ad   {a^{\dagger}}

\def\bv   {\overline {v}}
\def\bV   {\overline {V}}

\def\al   {\alpha}
\def\be   {\beta}
\def\bet  {\beta}

\def\HH {\hat H}

\def\HH {\hat H}

\def\CE {{\cal E}}
\def\CL {{\cal L}}

\def\CR {{\cal R}}

\def\CM {{\cal M}}

{\it}
{\it}
{\it}
\def\tr {\rm {tr}}{\it}
{\it}
{\it}
\def\be {\beta}
\def\bet {\beta}
\def\ga {\gamma}
\def\de {\delta}
\def\De {\Delta}

\def\sig {\sigma}

\def\om {\omega}
\def\Om {\Omega}

\usepackage{times}
\usepackage{graphicx}
\begin{document}
\title{ An efficient method to evaluate energy variances for extrapolation 
methods.}
\author{G. Puddu\\
       Dipartimento di Fisica dell'Universita' di Milano,\\
       Via Celoria 16, I-20133 Milano, Italy}
\maketitle
\begin {abstract}
         The energy variance extrapolation method consists in relating the approximate
         energies in many-body calculations to the corresponding energy variances and inferring
         eigenvalues by extrapolating to zero variance. The method needs a fast evaluation
         of the energy variances. For many-body methods that expand the nuclear wave functions in terms 
         of deformed Slater determinants, the best available method for the evaluation of energy 
         variances  scales with the sixth power of the 
         number
         of single-particle states. We propose a new method which depends on the number of single-particle
         orbits and the number of particles rather than the number of single-particle states.
         We discuss as an example the case of ${}^4He$ using the chiral N3LO interaction
         in a basis consisting up to $184$ single-particle states.
\par\noindent
{\bf{Pacs numbers}}: 21.60.De,$\,\,$  24.10.Cn,$\,\,$  27.10.+h
\vfill
\eject
\end{abstract}
\section{ Introduction.}
     Both the Monte Carlo shell model (MCSM) (refs. [1]-[3]) and the Hybrid Multi-Determinant 
     (HMD) method
     (refs. [4],[5]) approximate eigenstates of the nuclear Hamiltonian,
     although with a different parametrization and minimization method, with a linear combination of 
     Slater determinants as a variational ansatz.
     Several years ago (refs.[6]-[8]) a robust method has been proposed
     whereby the  error in the energy $E_{approx}-E_{exact}$ was shown to have a linear behavior 
     as a function of the energy variance for wave-functions sufficiently close to the exact ones.
     By extrapolating to zero variance one can then infer the exact (or almost exact) 
     energy eigenvalues. This idea was put forward for the first time in the context
     of shell model calculations where the energy variance is generated by the Lanczos diagonalization
     method. In ref. [6] a sequence of approximate shell model wave functions $|\psi>$ was
     obtained by truncation of number of possible excitations. Using the proportionality relation
     $<\psi|\HH|\psi>-E_{exact} \propto <\psi|\HH^2|\psi>-<\psi|\HH|\psi>^2$, valid if the approximate 
wave
     function is sufficiently close to the exact one, and extrapolating to zero variance  one can
     obtain the exact value of the energy, at least in principle.
     Within the Lanczos diagonalization method the variance is obtained without extra effort.
     For variational methods, although the extrapolation method is still valid, the evaluation
     of the energy variance is computationally very expensive. Note however that access to the full
     Hilbert space is not needed. Recently the extrapolation method has been
     revived within the Monte Carlo shell model method (refs.[9],[10]) and applied to ab-initio 
     calculations.
     The cost of the method is proportional to the sixth power of number of single-particle 
     states. The very same cost applies to the Hybrid Multi-Determinant method. Clearly for
     a systematic applicability of the method, the computational cost must be reduced.
     In this work  we propose an efficient method to evaluate the energy variance.
     More precisely we give an efficient recipe to evaluate $<\psi| \HH^2 | \phi>$ where $|\psi>$
     and $|\phi>$ are two Slater determinants. The method which will be described in detail in the 
     next section is based on a factorization of the density matrix and its computational costs 
     depends on the number of particles and of single-particle orbits $n,l,j$ rather than on the
     number of single-particle states $n,l,j,m$. In section 3 we discuss the application
     of the method to the case of ${}^4 He$ using the $N3LO$ interaction (ref.[11])
     in an harmonic oscillator
     basis of $6$, $7$  and $8$ major shells ($N_{ho}=5, 6, 7$  respectively).
     We restrict ourselves to $l\leq 5$ and the largest single-particle space consists of $184$
     single-particle states.
      Rather than work with the bare N3LO interaction we renormalize the interaction with
     two methods. In the first method we renormalize the interaction with a sharp 
     relative momentum cutoff $K_{max}=2.5 fm^{-1}$
     as done in $V_{low k}$ (ref. [12]), then we renormalize once more to a specified number of 
     harmonic oscillator shells using the  Lee-Suzuki renormalization procedure 
     (refs.[13],[14]). In the second method we use directly the Lee-Suzuki procedure on the N3LO
     interaction.
     The method we propose to evaluate the energy variance has been implemented on a personal computer
     with modest resources. 
\bigskip
\bigskip
\section{ Evaluation of the energy variance.}
\bigskip
     In both the MCSM and in the HMD methods the variational ansatz for the eigenstates is
$$
|\psi_N> = \sum_{\al=1}^N g_{\al N}|U_{\al,N}>,
\eqno(1)
$$
     Only for simplicity we omit the projector to good angular momentum and parity.
     The $|U_{\al_N}>$ are variational Slater determinants or,  more precisely, a product
     of a neutron and a proton Slater determinant.
     $N$ is the number of Slater determinants. If Slater determinants are added one by one 
     the cost of generating and optimizing $N$ Slater determinants scales as $N^2$.
     Therefore it is  computationally
     expensive to increase the accuracy in the evaluation of energies and observables after a 
     large number of Slater determinants. Therefore
     an efficient and theoretically robust extrapolation method is highly valuable.
     The coefficients $g_{\al,N}$ are determined by solving the generalized eigenvalue problem
$$
\sum_{\bet} <U_{\al,N}|\HH|U_{\bet,N}> g_{\bet,N} = E_N \sum_{\bet} <U_{\al,N}|U_{\bet,N}> g_{\bet,N},
\eqno(2)
$$
     $E_N$ being the approximate energy for a set of $N$ Slater determinants.
     We write the Hamiltonian as
$$
\HH={1\over 2} \sum \bv_{ij,kl} \ad_i\ad_j a_l a_k
\eqno(3)
$$
     where $\ad_i$ is the creation operator in the single-particle state $i$. For convenience
     we include the single-particle Hamiltonian in the anti-symmetric part of the two-body potential 
     $\overline v_{ij,kl}= {1 \over 2}(v_{ijkl}-v_{ijlk})$.
     The matrix elements of $\HH$ and of $\HH^2$ between two different Slater determinants
     $|U>$ and $|V>$ are given by
$$
{<V|\HH| U>\over <V|U>}=  \bv_{ijkl}\rho_{ki}\rho_{lj}
\eqno(4)
$$
     and
$$
{<V|\HH^2| U>\over <V|U>}= \big( \bv_{ijkl}\rho_{ki}\rho_{lj}\big) ^2 +
4 \tr \it{(\bV F \bV \rho)+\bv_{ijkl} F_{kp}F_{lq} \bv_{pqrs}\rho_{ri}\rho_{sj} }
\eqno(5)
$$
     where $\rho_{ki}= <V|\ad_i a_k|U>$ is the generalized density matrix, $F_{ik}=<V|a_i \ad_k|U>$
     and
$$
\bV_{jl}=\bv_{ijkl}\rho_{ki}
\eqno(6)
$$
     in eqs. (4)-(6), the sum over repeated indices is understood.
     Note that the indices of the density matrix refer only to identical particles, that is
     $\rho_{np}=0$.
\par\noindent
     The use of the anti-symmetric part of the Hamiltonian $\bv$
     reduces the number of terms since the exchange terms are opposite to the direct contributions.
     In eq. (5), the trace term is taken in the single-particle indices, i.e.
$$
tr (\bV F \bV \rho)= \bV_{ij} F_{jk} \bV_{kl}\rho_{li}
\eqno(7)
$$
     In ref.[9],apart a slight change in the notations, the last term in eq. (5) is recast as,  
$$ 
Q=\bv_{(ij), (kl) } (FF)_{(kl),(pq)} \bv_{(pq),(rs)} (\rho\rho)_{(rs),(ij)}
\eqno(8)
$$ 
     and it is treated as  two products of matrices of dimensions equal to $N_{sp}^2$, where $N_{sp}$
     is the number of single-particle states. Hence the computational cost scales as $N_{sp}^6$.
     
     We now proceed to modify these expressions in order to reduce the computational cost.
     The first two terms in eq.(5) pose no problem
     and we will focus entirely on the third term alone which we call $Q$. For readability we replace
     latin indices with numbers (e.g. $1\equiv(n_1,l_1,j_1,m_1)$ etc. of dimension $N_{sp}$)
\par
     The Slater determinants $|U>$ and $|V>$ are identified by the single-particle wave functions  
     $ U_{1,\al}$ and $V_{1,\al}$ with  and $\al=1,2,..A$ where $A$ is the number of 
     particles (we will treat explicitly neutrons and protons later).
     The expressions for $\rho$ and $F$ are given by
$$
\rho_{12}=\sum_{\al} U_{1\al}W_{\al,2}
\eqno(9a)
$$
      and
$$
F_{1 2}=\de_{1 2}-\rho_{1 2}
\eqno(9b)
$$
     where $\de$ is the Kronecker $\de$, and
$$
W_{\al,2}= \sum_{\bet}(V^{\dagger} U)^{-1}_{\al,\bet} V^{\dagger}_{\bet,2}
\eqno(10)
$$ 
     From eq.(9a) and eq.(10), one can see that $\rho^2=\rho$ and $\tr \rho=A$ and the matrix 
     multiplications
     in the expression for $\rho$ are of the type $(NA)\times(AA)\times(AN)$. 
     The rectangular matrices $U,W$ are stored  at the beginning of the calculations.
     The various $nn$, $pp$ and $np$ contributions give
$$
Q=Q^{(nn)}+Q^{(pp)}+4Q^{(np)}
\eqno(11)
$$
     the factor of $4$ comes from all possible exchanges between $n$ and $p$ indices, and
     $Q^{(np)}$ contains only terms like $<np|\bv|np>$, for sake of argument, in this order.
     Inserting eqs. (9a)-(10) in the last term of eq.(5) (the $Q$ term) and using
     eq.(9b), we obtain
$$
Q(tt')\equiv \bv_{1 2 3 4} F_{3 5}F_{4 6} \bv_{5 6 7 8}\rho_{7 1}\rho_{8 2}= 
C_2(tt')+C_{3a}(tt')+C_{3b}(tt')+C_4(tt')
\eqno(12)
$$ 
     for $(tt')=(nn),(pp),(np)$, with
$$
 C_2(tt')=\big(W_{\ga 1} W_{\de 2} \bv_{1234} \big) 
\big(\bv_{3 4 1' 2' }U_{1' \ga}U_{2' \de} \big)
\eqno(13a)
$$
$$
C_{3a}(tt')=-\big(W_{\ga 1} W_{\de 2} \bv_{1234}U_{4\be} \big) \big(W_{\be 4'} 
\bv_{34'1'2'}U_{1'\ga}U_{2'\de} \big)
\eqno(13b)
$$
$$
C_{3b}(tt')=-\big(W_{\ga 1} W_{\de 2} \bv_{1234}U_{3\al} \big) \big(W_{\al 3'} \bv_{3'4 
1'2'}U_{1'\ga}U_{2'\de} \big)
\eqno(13c)
$$
$$
C_4(tt') =(W_{\ga 1} W_{\de 2} \bv_{1234}U_{3\al}U_{4\be} )
( W_{\al 3'}W_{\be 4'} \bv_{3'4' 1'2'}U_{1'\ga}U_{2'\de} )
\eqno(13d)
$$
     Notice that in the case $(tt')=(np)$, odd numbers/letters represent neutrons 
     (e.g. $1 3 5,..,\al\ga\,..$) while even numbers/letters represent protons
     (e.g. $ 2,4,6,..\be\de,..$).
     Again the sum over repeated indices is understood.
     In eq.(13a)-(13d) we grouped the various terms so that matrix multiplications can
     be efficiently performed. Notice that 
$$
L_{\ga\de,34}= W_{\ga 1} W_{\de 2} \bv_{1234} 
\eqno(14a)
$$
     and
$$
R_{3 4 ,\ga\de}=\bv_{3 4 1'2'}U_{1'\ga}U_{2'\de}
\eqno(14b)
$$
     enter in all contributions and can be very efficiently evaluated since $\bv$ is very sparse.
     The various $C_2, C_3, C_4$ carry a label that identifies their 2-body, 3-body and 4-body 
     character. Each term contained between the brackets can be evaluated from the previous equations
     and using eqs.(14a),(14b).
     Eqs. (11)-(14) improve the computation of the $Q$ term, since some of the indices
     are particle indices, rather than single-particle ones. Essentially we make use of the fact
     that $\rho$ is a low rank matrix.
     In order to see how the computational cost scales with the particle number and with the number
     of single-particle states, consider for example the case $(tt')=(nn)$ and let $N_n$ be the number
     of neutrons and $N_v$ the number of non-zero matrix elements of $\bv$.
     It is easy to see that the computational cost for the evaluation once for all of the matrices
     $L$ and $R$ in eq.(14) scales as $N_v N_n^2$ and the evaluation of $C_2,C_3$ and $C_4$ 
     scale as $N_n^2 N_{sp}^2, N_n^3 N_{sp}^2$ and $N_n^4 N_{sp}$ respectively. Let us remark that
     in ab-initio calculations $N_{sp}>>N_n$. 
\par
     Despite this improvement, we prefer to use the angular 
     momentum coupled matrix elements of $\bv$. The reason is two-fold. First of all for large 
     single-particle
     spaces the number of uncoupled matrix elements of $\bv$ is very large. In some of the calculations described
     in the next section we have used a personal computer of 1Gyb of RAM,
     Second, instead of dealing
     as in eqs.(13a)-(13d) with single-particle indices, we prefer to deal with orbits (i.e. $nlj$).
\par\noindent
     Let us first change slightly the notations: let us label orbits by numbers, i.e. 
     $1\equiv(n_1,l_1,j_1)$ etc., while single-particle indices are now represented as
     $(1,m_1)$ etc., and, as before, greek letters count particles.
     It is then straightforward to arrive at the following results. 
     Define the angular momentum coupled quantities by Clebsh-Gordan coefficients
$$
R^{JM}_{3 4 \ga \de}= \bv^J_{3 4 1 2 } (UU)^{JM}_{1 2 \ga  \de}
\eqno(16)
$$
$$
L^{JM}_{\ga \de 3 4}= (WW)^{JM}_{\ga \de 1 2} \bv^J_{1 2 3 4}
\eqno(17)
$$
     with
$$
(WW)^{JM}_{\al\be 1 2}=\sum_{m_1m_2}<j_1 m_1 j_2 m_2 |JM> W_{\al,1 m_1} W_{\be,2 m_2}
\eqno(18)
$$
$$
(UU)^{JM}_{1 2 \al\be}=\sum_{m_1m_2}<j_1 m_1 j_2 m_2 |JM> U_{1 m_1,\al} U_{2 m_2,\be}
\eqno(19)
$$
     We stress again that in the case of the $(np)$ contribution, odd numbers/letters refer
     to neutrons and even numbers/letters refer to protons.
     Then the 2-body contribution to $Q$ is given by
$$
C_2(tt')=\sum_{JM} L^{JM}_{\al\be 1 2} R^{JM}_{1 2 \al \be}
\eqno(20)
$$
     where we have shown explicitly the sum over $JM$ only, and the sum over the remaining
     indices is implicit.
     All matrix multiplications have now small dimensions.
     For the contribution of 3-body type, define also
$$
W^{JM}_{\be 6 }(j_3 m_3)=\sum_{m_6} <j_3 m_3 j_6 m_6|JM> W_{\be, 6 m_6}
\eqno(21a)
$$
$$
U^{JM}_{4 \be }(j_3 m_3)=\sum_{m_4} <j_3 m_3 j_4 m_4|JM> U_{4 m_4, \be}
\eqno(21b)
$$
      and
$$
\CR_{\be,\ga\de}(3,m_3)= \sum_{JM, 6} W^{JM}_{\be,6}(j3,m3) R^{JM}_{3 6 \ga \de}
\eqno(22a)
$$
$$
\CL_{\ga \de,\be}(3,m_3)= \sum_{JM, 4} L^{JM}_{\ga\de, 3 4}U^{JM}_{4,\be}(j_3 m_3)
\eqno(22b)
$$
     Then
$$
C_{3a}(tt')=-\sum_{3 m_3}\CL_{\ga \de, \be}(3 m_3) \CR_{\be, \ga \de}(3 m_3)
\eqno(23)
$$
     In the case of $C_{3b}$ define
$$
U^{JM}_{3 \al}(j_4 m_4)=\sum_{m_3}<j_3 m_3 j_4 m_4 |JM> U_{3 m_3,\al}
\eqno(24a)
$$
$$
W^{JM}_{\al 5}(j_4 m_4)=\sum_{m_5}<j_5 m_5 j_4 m_4 |JM> W_{\al, 5 m_5}
\eqno(24b)
$$
     and
$$
\CL_{\ga\de,\al}(4,m_4)=\sum_{JM3} L^{JM}_{\ga \de,3 4} U^{JM}_{3 \al}(j_4 m_4)
\eqno(25a)
$$
$$
\CR_{\al,\ga\de}(4,m_4)=\sum_{5JM} W^{JM}_{\al,5}(j_4,m_4)R^{JM}_{5 4,\ga\de}
\eqno(25b)
$$
     Then
$$
C_{3b}(tt')=-\sum_{4,m_4}\CL_{\ga\de,\al} \CR_{\al,\ga\de}(4 m_4)
\eqno(26)
$$
    Note the difference in the implicitly summed indices in this expression compared with
    the ones in eq.(23).
     The contribution of 4-body type has the simple expression
$$
C_4(tt')= \CM_{\al \be, \ga \de} \CM_{\ga \de,\al\be}
\eqno(27)
$$
     with
$$
\CM_{\al \be, \ga \de}=\sum_{JM} (WW)^{JM}_{\al\be 1 2} R^{JM}_{1 2 \ga\de}
\eqno(28)
$$
     In the $np$ case the contributions $C_{3a}$ and $C_{3b}$ are not equal in general.
     They are equal in the case of $(tt')=(nn),(pp)$.
     Also it is worth to note that the four-fold sum in eq.(28) is over the particle indices only.
     As before, let us analyze how the computational cost scales the number of orbits and the number of 
     particles. Let us consider the case of the $nn$ contribution. The evaluation of eqs.(18) and (19)
     is straightforward because of the angular momentum selection rules. The evaluation of
     eqs.(16) and (17) scales as $N_{JM} N_v(JM) N_n^2$ where $N_{JM}$ is the number of the possible
     $JM$ values, $N_v(JM)$ is the number of non-zero matrix elements $ \bv^J_{1 2 3 4}$ (the total
     number of angular momentum coupled matrix elements$ \bv^J_{1 2 3 4}$ 
     can be a factor of $30$ smaller than the number of uncoupled matrix elements of $\bv$).
     Once eqs.(16) and (17) have been evaluated for all $JM$ values, the computational cost of $C_2$ 
     scales as $N_{JM} N_{orb}^2 N_n^2$ where $N_{orb}$ is the number of single-particle orbits.
     The evaluation of $C_3$ (cf. eq.(23)) scales as $ N_{sp} N_n^3$ in addition to the cost
     of evaluation of eqs.(22a) and (22b) which scales as $N_{JM} N_{sp} N_{orb} N_n^3$ (note that the 3-body
     term retains one power of $N_{sp}$. The computational cost of the evaluation of the matrix $\CM$ 
     scales as $N_{JM} N_{orb}^2 N_n^4$ and the evaluation of the 4-body contribution scales
     as $N_n^4$. Since the number of uncoupled matrix elements of the Hamiltonian can easily be of the
     order of $10^7$ we use only the angular momentum coupled representation.
\bigskip
\bigskip
\section{ A case study with ${}^4He$.}
\bigskip

      Let us apply the method derived in the previous section to ${}^4 He$. We use
      the N3LO interaction (ref. [11]) as the nucleon-nucleon potential. This potential has a smooth
      energy cutoff at $\Lambda=500 $MeV corresponding to a relative momentum
      $K_{max}=2.534 fm^{-1}$ The potential is however non-zero at higher relative momenta
      due to the smooth cutoff. In ref.[15], a rather large number of harmonic oscillator major
      shells was used in order to reach independence from the h.o. frequency and from the
      single-particle space, using the 'bare' potential. 
      In order to be able to work with smaller spaces we renormalize the interaction with
      two methods. In the first method we first renormalize the interaction much in the same way
      it is done to obtain $V_{lowk}$ interactions (cf. ref. [12]) to a sharp relative 
      momentum $K_{max}=2.5fm^{-1}$ (essentially to get rid of the high
      momentum tail) and then we renormalize once more to a specified number of major shells with the Lee-Suzuki method, in 
      order to reach some independence from the h.o. frequency with  reasonably small single-particle spaces.
      In the second method we apply directly the  Lee-Suzuki method to the N3LO interaction.
\par
      
      The rationale for the first choice is the following. For an interaction which has
      a sharp cutoff $K_{max}$ and for a system of radius $R$ the acceptable values of $\hbar\Om$
      are given by the inequality (ref.[15])
$$
\hbar^2 K_{max}^2/( m N_{ho})< \hbar \om < N_{ho} \hbar^2/(m R^2)
$$
      and in order to have a reasonable large interval in $\hbar \om $ we need large values
      of $N_{ho}$. This is correct if we use the bare interaction. It is hoped that, by 
      renormalizing
      the interaction one more time, we can reach the energies given by the bare interaction 
      having a specified $K_{max}$, with  smaller values of $N_{ho}$. The only problem
      with this argument is that the first renormalization to a sharp cutoff could give rise
      to induced many-body interactions. This is precisely the reason why we took
      a large value of $K_{max}$, that is, to minimize the induced many-body interactions. In ref.
      [17], it was shown that in order to minimize the effect of induced many-body interactions,
      $K_{max}$ should be as large as possible ($K_{max}\simeq 5 fm^{-1}$), this result 
      has been obtained using the CD-Bonn interaction which contains larger energy scales than
      N3LO. In any case, if we use $V_{low k}$ interactions with a large cutoff, another renormalization
      step is nearly unavoidable, unless we are willing to work with very large single-particle spaces.  
\par
      We have used the HMD method of type-a (described in ref. [16])
      to generate a sequence of wave-functions $|\psi_n>$ consisting of a $n$ Slater determinants
      (up to a maximum of $100$).
      We have used a projector
      to good parity and good z-component of the angular momentum.
      We considered several values of the harmonic oscillator frequency namely $\hbar\Om=(21,24,27,30,33,36) MeV $
      and $N_{ho}=5,6$. For $N_{ho}=7$ we considered only $\hbar\Om=(27,30)MeV$.
      Only states with $l<6$ have been included in 
      the single-particle basis.
      The largest number of single-particle states is $184$  and the corresponding number of orbits is $32$ for $N_{ho}=7$.
\par
\renewcommand{\baselinestretch}{1}
\begin{figure}
\centering
\includegraphics[width=10.0cm,height=10.0cm,angle=0]{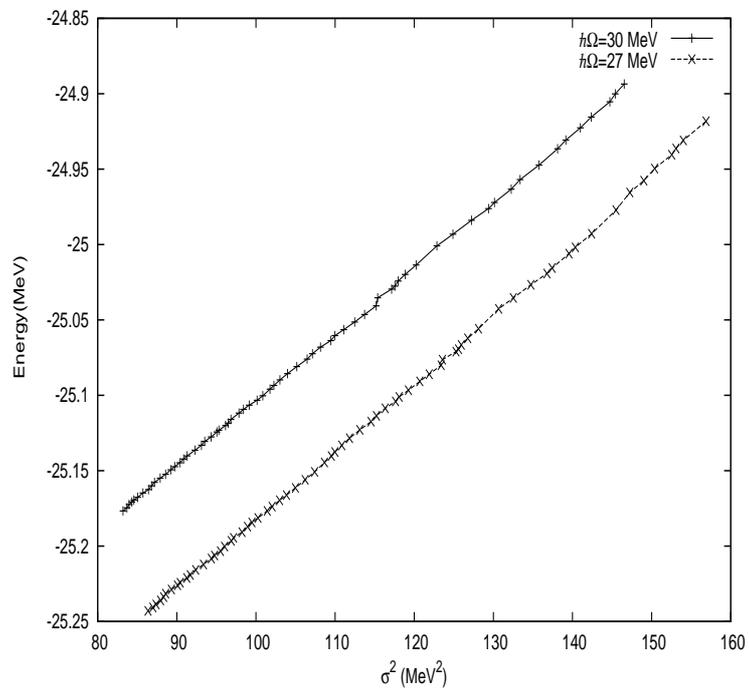}
\caption{Energy-variance plot for $N_{ho}=7$.}
\end{figure}
\renewcommand{\baselinestretch}{2}
      Among the several variants of the energy vs. variance extrapolation methods 
      we considered the first one, used in ref.[6], where $E_n= <\psi_n|\HH|\psi_n>$ is expanded as a 
      function of the dimensionless variance defined as
$$
\De E_n= {\sig^2\over <\psi_n|\HH|\psi_n>^2}\equiv {<\psi_n|\HH^2|\psi_n>-<\psi_n|\HH|\psi_n>^2\over 
<\psi_n|\HH|\psi_n>^2}
$$
      where $|\psi_n>$ is the wave-function obtained with the first $n$ Slater 
      determinants,
      and fitted with a linear function
      $ E_n= a +b\De E_n$. In the limit $\De E_n=0$,  $a$ is the extrapolated ground-state energy.
\par
\renewcommand{\baselinestretch}{1}
\begin{figure}
\centering
\includegraphics[width=10.0cm,height=10.0cm,angle=0]{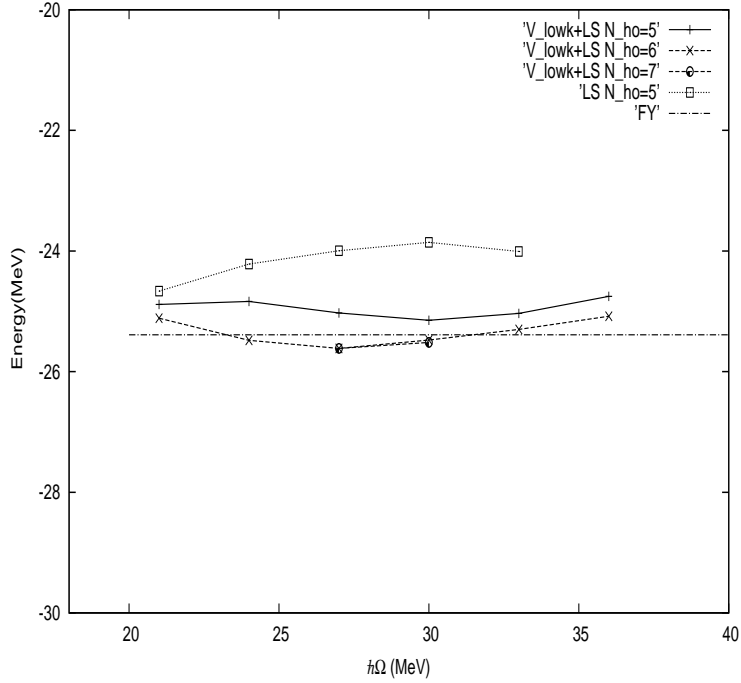}
\caption{Energies as a function of $\hbar\Om$.}
\end{figure}
\renewcommand{\baselinestretch}{2}

      For the purpose of the extrapolation we discard low values of $n$ as done in ref. [9].
      As a measure of the 
      soundness of the linear fit we consider the quantity 
$$
\CE_{n_0}= \sqrt{\sum_{n>n_0} (E_n - a  -b\De E_n )^2}
\eqno(29)
$$
      where the first $n_0$ many-body wave functions have been discarded, since they are
      a poor approximation to the exact eigenstate.
      The value of $n_0$ is selected so that $\CE_{n0}$ is of the order of $10 KeV$ or smaller.
      In fig. 1 we show a few plots of the energy as a function of the variance $\sig^2$.
      Typically variances for this nucleus and this interaction are much larger
       than the ones shown in ref. [9]. This is not very surprising since in our case
      the spectrum extends over very large values of the energy, while in the case study 
      of ref. [9], where only one and two major shells were considered, the spectrum
      is much more compressed.
      In fig. 2 we show our results of the calculation and the the exact Faddeev-Yakubovski
      result. In fig. 2 we also show the results obtained using directly the Lee-Suzuki method
      for $N_{ho}=5$. Interestingly enough, the double renormalization method performs better
      than the single L-S renormalization which shows a too marked variation as a function
      of the harmonic oscillator frequency. Note also the results for $N_{ho}=7$ are very close
      to the ones for $N_{ho}=6$.  
\par
      In conclusion, in this work we have implemented an efficient method to evaluate
      energy variances for variational many-body calculations which are needed in order
      to apply the energy variance extrapolation method. This method avoids the need to  
      determine a very large number of Slater determinants in order to improve the accuracy
      of observables. Since the computational cost of the evaluation of $N$ Slater determinants
      scales as $N^2$, the method is a useful tool to decrease the cost of variational 
      many-body calculations.

\vfill
\eject

\begin{thebibliography}
\bigskip
\bibitem{1}
  M.Honma, T.Mizusaki, and T.Otsuka, Phys. Rev. Lett. 75, 1284 (1995).
\bibitem{2}
  T.Otsuka, M.Honma, and T.Mizusaki, Phys. Rev. Lett. 81,1588 (1998).
\bibitem{3}
  T. Otsuka, M. Honma, T. Mizusaki, N. Shimizu, and Y. Utsuno.\\
  Prog. Part. Nucl. Phys. 47, 319 (2001).
\bibitem{4}
    G.Puddu. J. Phys. G: Nucl. Part. Phys. {\bf 32},321 (2006).
\bibitem{5}
    G.Puddu. Eur. Phys. J. A 34, 413 (2007).
\bibitem{6}
T. Mizusaki and M. Imada. Phys. Rev. C 65, 064319(2002).
\bibitem{7}
T. Mizusaki and M. Imada. Phys. Rev. C  67, 041301 (2003).
\bibitem{8}
 T. Mizusaki, Phys. Rev. C 70, 044316 (2004).
\bibitem{9}
N.Shimizu, Y.Utsuno, T.Mizusaki, T.Otsuka, T.Abe, and M.Honma.\\
   Phys. Rev. C 82,061305(2010).
\bibitem{10}
   T.Abe, P.Maris, T.Otsuka, N.Shimizu, Y.Utsuno and J.P.Vary.\\
   arXiv: 1107.1784 [nucl-th].
\bibitem{11}
 D. R. Entem and R. Machleidt, Phys. Rev. C 68, 041001(2003).
\bibitem{12}
  J.D. Holt, T.T.S.Kuo, and G.E.Brown, Phys. Rev. C 69, 034428(2004).
\bibitem{13}
 K.Suzuki and S.Y.Lee. Prog. Theor. Phys. 64, 209,(1980).
\bibitem{14}
 S.Fujii, R.Okamoto, and K.Suzuki. Phys. Rev. C 69, 034328 (2004).
\bibitem{15}
  G.Hagen, T.Papenbrock, D.J.Dean,and M.Hjorth-Jensen.\\
  Phys. Rev. C 82, 034330 (2010).
\bibitem{16}
 G.Puddu. Eur. Phys. J. A 45, 233(2010).
\bibitem{17}
    S.Fujii, E.Epelbaum, H.Kamada, R.Okamoto, K.Suzuki, \\
   and W. Glockle.  Phys. Rev. C 70,024003(2004).
\end{thebibliography}
\end{document}